\documentclass[prd,twocolumn,showpacs,preprintnumbers,amsmath,amssymb]{revtex4}

\usepackage{graphicx}
\usepackage{dcolumn}
\usepackage{bm}

\def\be{\begin{eqnarray}}
\def\ee{\end{eqnarray}}
\def\bea{\begin{eqnarray}}
\def\eea{\end{eqnarray}}
\def\kT{{\bf k}_\perp}
\def\pT{{\bf p}_\perp}
\def\PT{{\bf P}_\perp}
\def\sT{{\bf s}_\perp}
\def\yT{{\bf y}_\perp}
\def\zT{{\bf z}_\perp}
\def\bT{{\bf b}_\perp}
\def\IT{{\bf I}_\perp}
\def\RT{{\bf R}_\perp}
\def\DT{{\bf \Delta}_\perp}
\def\D2{{\bf \Delta}_\perp^2}
\def\0T{{\bf 0}_\perp}
\begin{document}


\title{Chromodynamic Lensing and Transverse Single Spin Asymmetries}

\author{Matthias Burkardt}
 \affiliation{Department of Physics, New Mexico State University,
Las Cruces, NM 88003-0001, U.S.A.}

\date{\today}

\begin{abstract}
We illustrate how an axial asymmetry in impact parameter 
dependent parton distributions can give rise to an axial
asymmetry for the transverse momentum of the leading quark
in the photo-production of hadrons. The effect is related
to the asymmetry originating from the Wilson-line phase factor
in gauge invariant Sivers distributions.
The single spin asymmetry arising from the asymmetry of the 
impact parameter dependent parton distributions is shown to 
exhibit a pure $\sin \phi$ dependence.
\end{abstract}

\maketitle
\section{Introduction}
Many semi-inclusive hadron production experiments show surprisingly 
large transverse polarizations or single spin
asymmetries (SSA) \cite{lambda}.
Moreover, the signs of these polarizations are usually
not dependent on the energy.
This very stable polarization pattern suggests that
there is a simple mechanism that gives rise to 
these polarization effects.

Recently, it has become clear that phase factors due to
final state interactions (FSI) of the struck quark
play a crucial role for the leading twist single spin
asymmetry (SSA) in semi-inclusive DIS \cite{hwang}.
Likewise, the initial state interactions (ISI) 
of the annihilating antiquark and the spectator quarks
is believed to give rise to SSA in the corresponding
Drell-Yan process.

According to BHS \cite{hwang}, 
the SSA depends on the interference of 
different amplitudes arising from the hadron's wavefunction and is
distinct from probabilistic measures of the target such as 
transversity.

As has been emphasized in Ref. \cite{collins}, this mechanism 
is also consistent with the Sivers mechanism.
In Collins' treatment the FSI of the struck quark are 
incorporated into Wilson line path-ordered exponentials 
(see also Ref. \cite{ji}).
The ISI and FSI of the struck quark with the gluon field
produce T-odd spin correlations (BHS), which is why a Sivers
asymmetry is allowed.

However, what is not clear from these treatments is why the
resulting SSAs are so large and
and exhibit such stable patterns.
The main purpose of this paper is to investigate
whether one can understand, in a more physical picture,
the mechanism associated with these phase factors
that gives rise to such large SSAs.

\section{Initial (final) state interactions and transverse spin
asymmetries}
Ref. \cite{collins} explains how ISI and
FSI allow the existence of T-odd parton distribution functions.
Formally the FSI (ISI) can be incorporated into $\kT $
dependent parton distribution functions (PDFs) by 
introducing a gauge string from each quark field
operator to infinity 
\be
P(x,\kT,\sT) &=& \int \frac{dy^- d^2\yT }{16\pi^3}
e^{-ixp^+y^-+i\kT \cdot \yT } \label{eq:P}\\
& &\!\!\times
\left\langle p \left|\bar{q}(0,y^-,\yT) W^\dagger_{y\infty}
\gamma^+ W_{0\infty} q(0)\right|p\right\rangle .
\nonumber
\ee
We use light-front (LF) coordinates are defined as:
$y^\mu =(y^+,y^-,\yT)$, with $y^\pm = (y^0\pm y^3)/\sqrt{2}$.

$W_{y\infty}=P\exp \left(-ig\int_{y^-}^\infty dz^- A^+(y^+,z^-,\yT)
\right)$ indicates a path ordered Wilson-line operator 
going out from the point $y$ to infinity.
The specific choice of path in Ref. \cite{collins} reflects the
FSI (ISI) of the active quark in an eikonal approximation.

The complex phase in Eq. ({\ref{eq:P}) is reversed under 
time-reversal
and therefore T-odd PDFs may exist \cite{collins},
which is why a nonzero Sivers asymmetry 
\cite{sivers} is possible.
However, given the fact that the asymmetry in 
$\kT$ hinges on
a complex phase that depends on gluon-fields, it remains a
puzzle why the resulting polarizations are so large and often
not very
sensitive to parameters like the energy or the momentum transfer.

Likewise, from the point of view of light-cone wave functions,
spin asymmetries arise from the phase difference between two
amplitudes coupling the proton target with
$J_p^z=\pm \frac{1}{2}$ to the same final state 
\cite{hwang}. 
The main purpose of this paper is to investigate, in a semi-classical
picture, how this phase translates into stable and
large SSA. 

\label{sec:main}
The physics of the transverse asymmetry can be best seen by
focusing on the mean transverse momentum
\be
\PT (x,\sT) &\equiv&
\int d^2\kT P(x,\kT,\sT)\, \kT
\label{eq:Pp}\\ &=&
i\int \frac{dy^- }{4\pi}
e^{-ixp^+y^-}\nonumber\\
& &\!\!\!\!\!\!\!\!\!\!\!\!\!\!\!\!\!\!\times
\left\langle p \left|\frac{\partial}{\partial_{\yT}}
\left.\bar{q}(0,y^-,\yT) W^\dagger_{y\infty}\right|_{\yT=0}
\gamma^+ W_{0\infty} q(0)\right|p\right\rangle .
\nonumber
\ee
The $\perp$ derivative in Eq. (\ref{eq:Pp}) can act both on the
quark field operator as well as on the gluon string.

Before proceeding further we would like to switch to light-front
gauge $A^+=0$. Already in an abelian theory 
\be 
a(\zT)\equiv \int_{-\infty}^\infty dz^- 
A^+(z^-,\zT)
\ee
is a gauge invariant quantity, 
it not entirely possible to
accomplish this (In a nonabelian theory, the gauge
invariant quantity is $tr\left(W_{-\infty,\infty}\right)$,
and the argument is similar).
In a box of length $L$ with periodic boundary conditions in the $z^-$-direction, 
the closest one can get to LF gauge is
$A^+(z^-,\zT)=\frac{1}{L}a(\zT)$
\cite{zero}. If we regulate these zero-modes by working in such a
box, using a gauge $A^+=const$ and finally taking 
$L\rightarrow \infty$, what we find is that (see also Appendix
\ref{sec:details})
\be
\int_{y^-,\yT}^{\infty,\yT} 
dz^- A^+(z^-,\yT)
&\longrightarrow& 
\frac{1}{2}\int_{-\infty,\yT}^{\infty,\yT} 
dz^- A^+(z^-,\yT) \nonumber\\
&=& \frac{1}{2} a(\zT) 
\ee
and\footnote{Note that the path-ordering of the gauge
string becomes unnecessary for a constant gauge
field.}  
\be
W_{y^-,\yT,\infty}\rightarrow e^{-\frac{i}{2}g a(\yT)}
\equiv w(\yT) 
\label{eq:w}
\ee
becomes independent of $y^-$.

In this work, we conjecture that even though a 
strict light-cone
gauge cannot be achieved, one can nevertheless 
summarize the mean 
effects of the $A^+$ component in its zero-mode 
(\ref{eq:w}). Likewise, any light-like
gauge string that does not extend to infinity 
becomes trivial
\be
W_{y^-\!,{\bf y}_T;z^-\!,{\bf y}_T}\rightarrow 1,
\ee
which is why the gauge string can be omitted in the
light-like correlations relevant for inclusive DIS.

In such an ``almost-LF gauge'', one thus finds that
\be
\PT(x,\sT)
&=& 
i\int \frac{dy^- }{4\pi} e^{-ixp^+y^-}\label{eq:Pw}\\
& &\!\!\!\!\!\!\!\!\!\!\!\!\!\!\!\!\!\!\!\!\!\!\!\!\!\!\!\!\!\!\!\!\!\!
\times
\left\langle p \left|\frac{\partial}{\partial_{\yT}}
\left.\bar{q}(0,y^-,\yT) w^\dagger(\yT)\right|_{\yT=0}
\gamma^+ w(\0T) \psi(0)\right|p\right\rangle .
\nonumber
\ee
In the term where the derivative acts on the quark field the
gauge string contribution disappears
\be
& &\!\!\!\!\!\!\!\!\!\!\!\!\!\!\!\!\!\!\!\!
\left\langle p \left|\partial_T\left(\bar{q}(0,y^-,\0T) 
\right)
w^\dagger (\0T)
\gamma^+ w(\0T) \psi(0)\right|p\right\rangle \nonumber\\
&=& 
\left\langle p \left|\partial_T\left(\bar{q}(0,y^-,\0T) 
\right)
\gamma^+ \psi(0)\right|p\right\rangle 
.
\label{eq:quark}
\ee
However, without the gauge string pointing to infinity,
\be
\left\langle p \left|\partial_T\left(\bar{q}(0,y^-,\0T) 
\right)
\gamma^+  \psi(0)\right|p\right\rangle =0,
\ee
due to time reversal invariance.

Hence the only non-vanishing contribution in Eq. (\ref{eq:Pw})
arises when the derivative acts on the gauge string. Upon introducing
\be
\IT (\0T) &\equiv& \left.
i\partial_{\yT}w^\dagger (\yT)\right|_{\yT=0} w(\0T)
\\
&=& -\frac{g}{2}\int_{-\infty}^\infty dz^-
\partial_{\yT} A^+(z^-,\0T) = -\frac{g}{2}\partial_{\yT}a(\0T) 
\nonumber
\ee
we can thus rewrite Eq. (\ref{eq:Pw}) in a very compact form
\be
\PT(x,{\bf s_\perp}) &=&
\int \frac{dy^- }{4\pi}
e^{-ixp^+y^-}\label{eq:Pi}\\
& &\times
\left\langle p \left|\bar{q}(0,y^-,\0T) {\bf I}_\perp(\0T)
\gamma^+ q(0)\right|p\right\rangle ,
\nonumber
\ee
which has a very physical interpretation as
the correlation between the transverse quark position and
the transverse impulse $\IT (\0T)$
acquired by the active quark as it escapes to infinity.
This interpretation becomes even more transparent after switching
to an impact parameter representation \cite{soper,me:1st,diehl,ijmpa}
\be
\left| p^+,\RT ,s \right\rangle \equiv {\cal N}
\int \frac{d^2{\bf p}_\perp}{2\pi}
\left| p^+,\pT,s \right\rangle e^{-i\pT\cdot\RT},
\ee
where ${\cal N}$ is some normalization.
Using the fact that the correlator in Eq. (\ref{eq:Pw})
does not change the transverse center of momentum 
(i.e. it is diagonal in $\RT $) this yields
\be
\PT(x,\sT) &=&\int d^2\RT \int \frac{dy^- }{4\pi} e^{-ixp^+y^-}
\label{eq:PR}\\
& &\!\!\!\!\!\!\!\!\!\!\!\!\!\times
\left\langle p^+, \RT \left|\bar{q}(y^-,\0T) \IT (\0T)
\gamma^+ q(0)\right|p^+,\RT \right\rangle \nonumber\\
&=&\int d^2\RT \int \frac{dy^- }{4\pi} e^{-ixp^+y^-}\nonumber\\
& &\!\!\!\!\!\!\!\!\!\!\!\!\!\!\!\!\!\!\!\!\!\!\!\!\!\!\!\!\!\times
\left\langle p^+\!, \0T \left|\bar{q}(y^-\!,-\RT) \IT(-\RT)
\gamma^+ q(0^-\!,-\RT)\right|p^+\!, \0T \right\rangle,\nonumber\\
&=&\int d^2\bT \int \frac{dy^- }{4\pi} e^{-ixp^+y^-}\nonumber\\
& &\!\!\!\!\!\!\!\!\!\!\!\!\!\!\!\!\!\!\!\!\!\!\!\times
\left\langle p^+, \0T \left|\bar{q}(y^-,\bT) \IT(\bT)
\gamma^+ q(0^-,\bT)\right|p^+, \0T \right\rangle,\nonumber
\ee
where we used translational invariance and then
substituted the (dummy)-integration 
variable $\RT$ by the 
integration variable $-\bT$. We use
$\left\langle p^+, \0T \right|$ as
a shorthand notation for $\left\langle p^+, \RT=\0T \right|$. 
If we compare this result with the 
impact parameter dependent parton distributions 
\cite{soper,me:1st,diehl,ijmpa}
\be
q(x,\bT) &=&\!
 \int \frac{dy^- }{4\pi} e^{-ixp^+y^-}\label{eq:ipd}\\
& &\!\!\!\!\!\!\!\!\!\!\!\!\!\!\times
\left\langle p^+, \0T \left|\bar{q}(y^-,\bT ) 
\gamma^+ q(0^-,\bT)\right|p^+,\0T \right\rangle,
\nonumber
\ee
we realize that Eqs. (\ref{eq:PR}) and (\ref{eq:ipd})
differ only by the presence of the operator $\IT(\bT)$ and an 
integration over the impact parameter $d^2\bT$.
What we have thus accomplished is to express the
mean transverse momentum of the outgoing quarks
in terms of a correlation between impact parameter
dependent PDFs and the transverse impulse 
$\IT(\bT)$ as a function of impact parameter.

The physical interpretation
of our result (\ref{eq:PR}) is thus evident:
$\IT(\bT)$ is the net transverse impulse that 
a quark at $\perp$ position $\bT$ 
receives on its way out.
The mean transverse momentum of the outgoing quark 
can be obtained by correlating the impact 
parameter dependent parton distribution
with the impact parameter dependent impulse for an 
outgoing quark.

For a transversely polarized target, the
impact parameter dependent parton distribution is
not axially symmetric \cite{me}. Therefore, even
if one assumes to lowest order that $\IT$ was 
axially symmetric (which it actually does not have 
to be), the momentum distribution of the
outgoing quark still exhibits an axial asymmetry.
And the asymmetry arises from correlating
the impact parameter dependent PDFs with the
impact parameter dependence of the impulse due
to the FSI.
This is one of the main results of this paper and 
provides a mathematical foundation for the 
heuristic model for SSAs that was advocated in 
Ref. \cite{me}:

For a transversely polarized target, the impact parameter dependent 
parton distribution $q(x,\bT)$ is no longer axially symmetric. 
In Ref. \cite{me} it was suggested that the final state interactions
deflect the outgoing quark in such a way that it receives a
transverse momentum that is directed toward the center of the 
target. As a result, the final state interactions 
thus translate the axial asymmetry in impact 
parameter space into an axial asymmetry in the 
transverse momentum. Eq. (\ref{eq:PR}) demonstrates 
that this very physical picture for SSA can in fact 
be related to the Wilson phase contribution 
discussed in Refs. \cite{hwang,collins}.

\section{Simple Models for the Final State Interactions}

In order to illustrate the implications of our 
results we will in the following adapt a potential 
model and treat the gluon vector potential as if it 
was abelian
\be
gA^0({\vec r}) &=& V(r) \nonumber\\
g{\vec A}({\vec r}) &=& 0.
\label{eq:potential}
\ee
These are clearly drastic approximations, but what 
we have in mind is not an exact treatment of the 
problem but rather a qualitative illustration of 
the physics that is connected with these phase 
factors.

Note that the vector potential in Eq. 
(\ref{eq:potential}) does not satisfy $A^+=const.$.
In principle, one could transform the above ansatz
into such a gauge. 
However, this is not necessary since $\IT$ is 
gauge invariant if we compactify space and we can 
evaluate $\IT$ it in any gauge.

The specific models that we consider are a 
logarithmic potential,
linear, as well as quadratic confinement.
\be
V_a(r) &=& c \ln \frac{r}{r_0} \nonumber\\
V_b(r) &=& \sigma r \Theta(R-r) + \sigma R \Theta(r-R)
\nonumber\\
V_c(r) &=& \frac{K}{2}r^2\Theta(r-R) +\frac{K}{2}R^2\Theta(R-r)
.
\ee
In the cases $b$ and $c$ we have to introduce a long distance
cutoff in order to avoid infrared divergences in
$\int dz {\bf \partial}_\perp A^0$. The physical mechanism for such
a cutoff is provided by pair creation when the active quark has 
separated far enough from the target.
Parameters: $\sigma=1\frac{GeV}{fm}$, $K=1.4 \frac{GeV}{fm^2}$, and
$c=0.3\, GeV$.
The cutoff radius is somewhat arbitrary and we chose a value
of $R=1\,fm$.

The lensing function that assigns a mean transverse momentum
for each impact parameter
\be 
\IT (\bT) = -{\bf \nabla}_\perp \frac{1}{2}\int_{-\infty}^\infty
dz V(\sqrt{\bT^2 + z^2})
\ee
in these three models is given by
\be
\IT^a (\bT) &=& -\frac{c\pi}{2} \frac{\bT}{\left|\bT\right|}
\nonumber\\
\IT^b(\bT) &=& -\frac{\sigma \bT}{2} \ln \left(
\frac{R+\sqrt{R^2-\bT^2}}{R-\sqrt{R^2-\bT^2}}\right)\Theta(R^2-\bT^2)
\nonumber\\
\IT^c (\bT) &=& -2k \bT\sqrt{R^2-\bT^2}\Theta(R^2-\bT^2) .
\label{eq:models}
\ee
\begin{figure}
\unitlength1.cm
\begin{picture}(10,15.5)(2,2.7)
\includegraphics{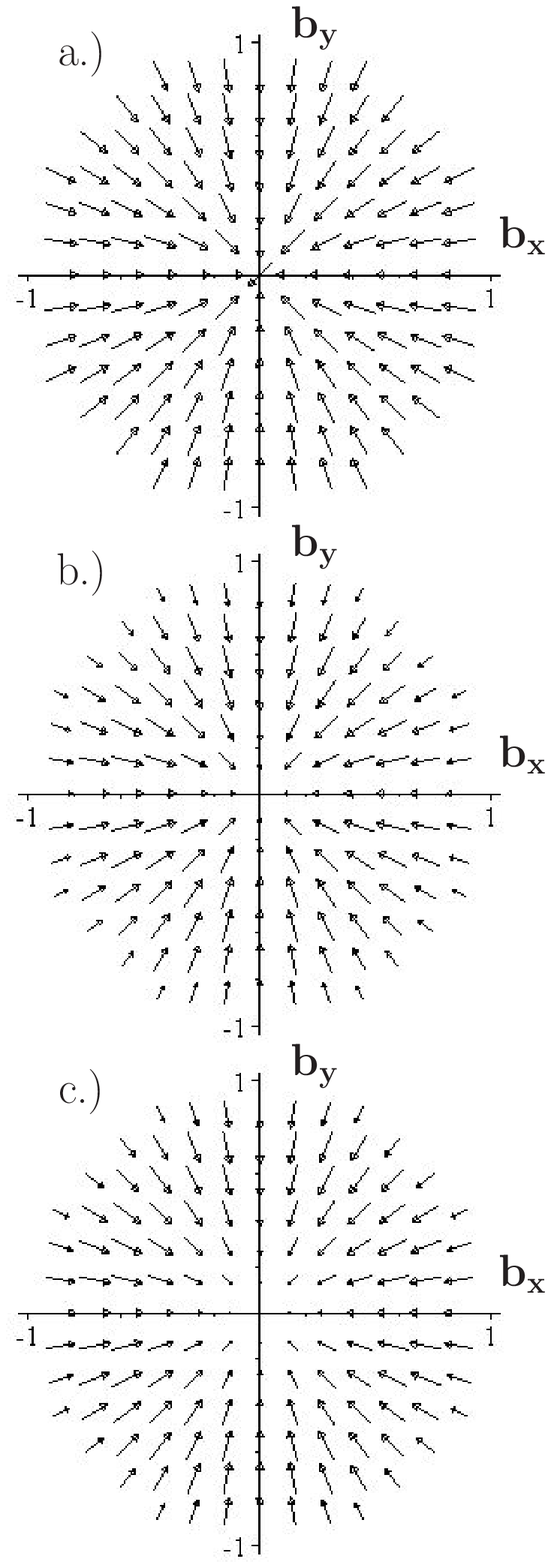}
\end{picture}
\caption{``Lensing function'' $\IT$ for $|\bT|<1\,fm$ in
the three models (\ref{eq:models}) for the quark potential.
The vector field represents the mean transverse momentum
that the ejected quark acquires when it is knocked out at
transverse position (relative to the center of momentum)
$\bT$.}
\label{fig:fpanel}
\end{figure}  
It is not surprising to find in all 3 models that 
the momentum is directed opposite in direction to 
the original transverse position $\bT$ since the 
underlying potentials are all attractive. In fact, this
should be a model-independent feature.

From Fig. \ref{fig:fpanel} one can also see that,
although there are some differences in the details,
the transverse momenta generated by a quark being
ejected through these momenta are all of the same 
order of magnitude $0.3-0.5\,GeV$ (Fig. \ref{fig:itbt})
\begin{figure}
\unitlength1.cm
\begin{picture}(10,6.5)(1.6,12.)
\includegraphics{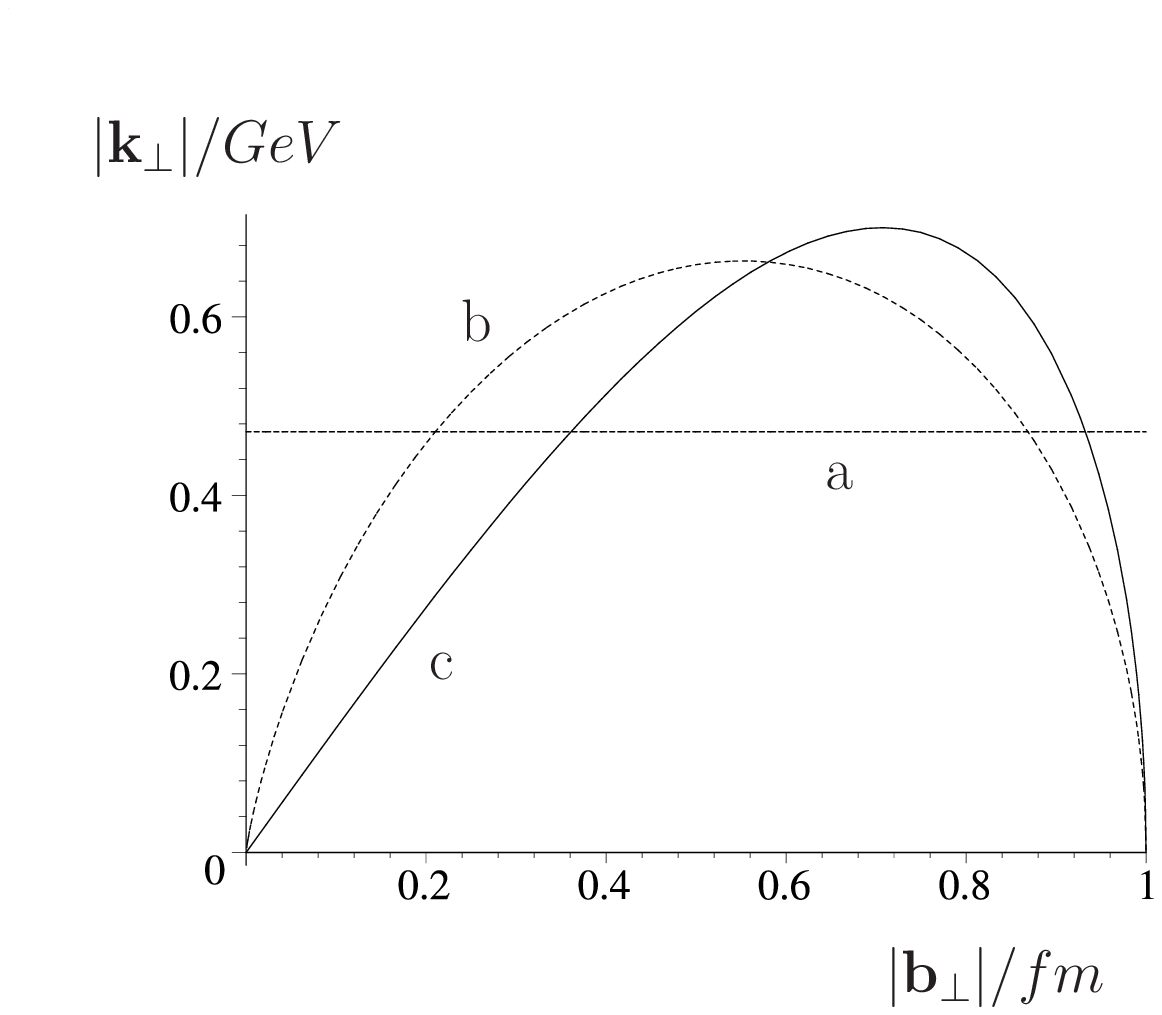}
\end{picture}
\caption{Transverse momenta resulting from
the three models (\ref{eq:models}) for the quark potential
as a function of the impact parameter $\bT$.}
\label{fig:itbt}
\end{figure}  
The only significant differences arises for very
small $\bT$ due to the very different short distance
behavior when one compares logarithmic, linear,
and quadratic potentials.

If the target is polarized in the x-direction in
the infinite momentum frame, the unpolarized
impact parameter dependent PDFs for flavor $q$
reads \cite{ijmpa}
\bea
q(x,\bT)&=&
\int\!\!\frac{d^2\DT}{(2\pi)^2}
e^{-i\DT\cdot\bT}\label{eq:qX}\\
& &\times\left[ H_q(x,0,\!-\D2) 
+ \frac{i\Delta_y}{2M} E_q(x,0,\!-\D2)\right]
\nonumber\\
&=& q(x,\bT) - \frac{1}{2M}\frac{\partial}
{\partial b_y} {\cal E}_q(x,\bT),
\nonumber
\eea
where we denoted ${\cal E}_q$ the Fourier transform
of $E_q$, i.e.
\be
{\cal E}_q(x,\bT)\equiv
\int \frac{d^2\DT}{(2\pi)^2}e^{-i\DT\cdot\bT}
E_q(x,0,-\D2).
\ee
Several comments are in order: first, Eq. 
(\ref{eq:qX}) applies to a nucleon that is 
polarized in the $x$ direction in the 
infinite momentum frame.
If one boosts this result into the rest frame,
relativistic corrections from the boost arise and
one needs to replace $E_q$ by $E_q+H_q$ in the
term that described the asymmetry. Secondly, it
should be emphasized that the axial
asymmetry in
Eq. (\ref{eq:qX}) is described by the 
$y-$derivative of an axially symmetric function,
i.e. the angular dependence is proportional to
$\sin (\phi)$, where $\phi$ is the angle relative
to the (transverse) spin direction. No higher
moments (e.g. $\sin (2\phi)$) are present.

The impact parameter dependent PDFs $H_q(x,0,-\D2)$ 
and $E_q(x,0,-\D2)$ are not known yet. In order to
proceed, we thus adopt the simple model from 
\cite{ijmpa}, where
\be
H_q(x,0,t) &=& q(x) e^{at(1-x)\ln\frac{1}{x}}
\nonumber\\
E_d(x,0,t) &=& \kappa_d H_d(x,0,t)\nonumber\\
E_u(x,0,t) &=& \frac{1}{2}\kappa_u H_u(x,0,t),
\label{eq:model}
\ee
where $\kappa_q$ is the contribution from quark
flavor $q$ to the anomalous magnetic moment of the
proton. The factor $\frac{1}{2}$ for down quarks
reflects the fact that $\int dx H_u(x,0,t)=2$.
Typical results for the resulting $\perp$ distortion 
in the model from Ref. \cite{ijmpa} are shown 
in Fig. \ref{fig:distort}.

\begin{figure}
\unitlength1.cm
\begin{picture}(10,10.)(2,7.7)
\includegraphics{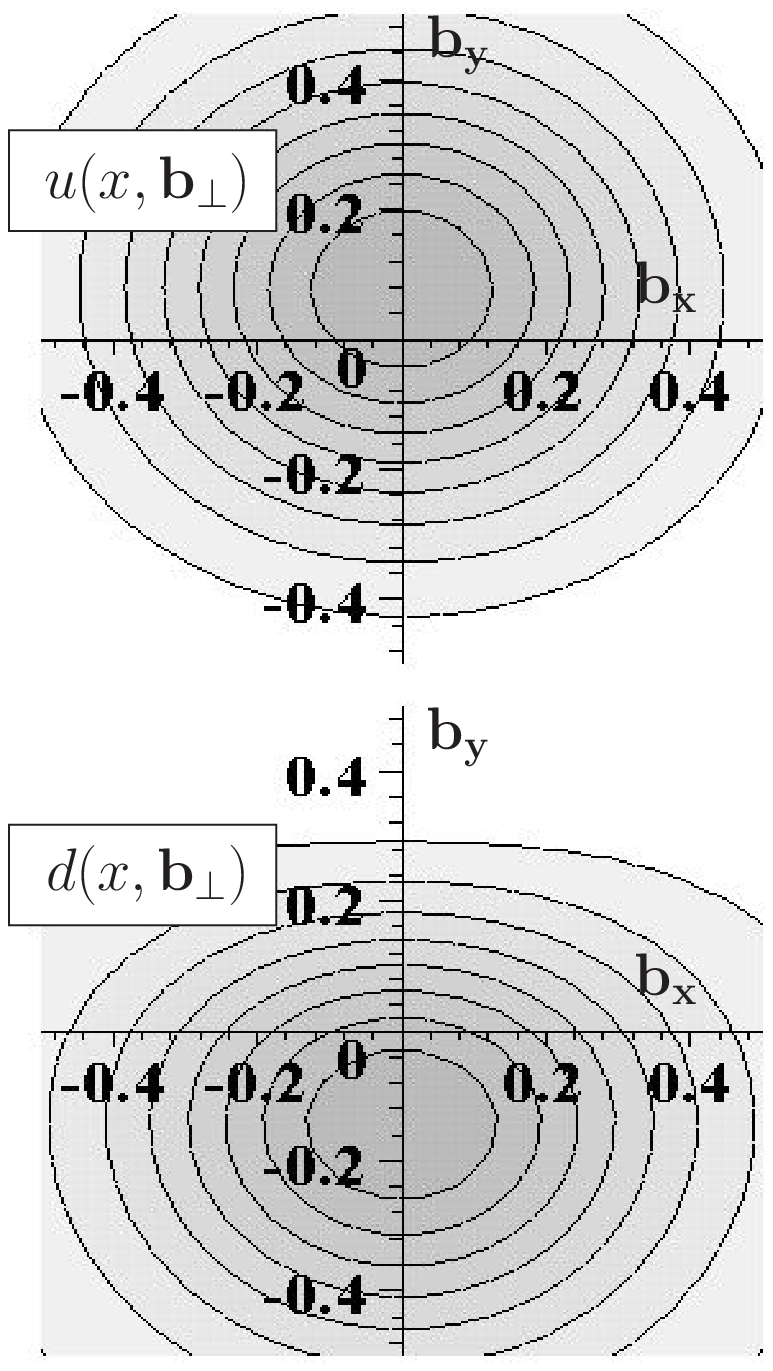}
\end{picture}
\caption{Distribution of $u$ and $d$ quarks in the
$\perp$ plane ($x_{Bj}=0.3$ is fixed) for a nucleon that is polarized
in the $x$ direction (\ref{eq:qX}) in the model from
Ref. \cite{ijmpa} (\ref{eq:model}).}
\label{fig:distort}
\end{figure}  
The naive model incorporates a number of different
features know about $GPDs$: in the forward limit
they reduce to the usual PDFs, at small $x$ the
$\perp$ size of the proton grows like 
$\ln\frac{1}{x}$, and this ansatz also satisfies 
duality.
Although some features, e.g. 
$\frac{QF_2}{F_1}\sim const.$ are not correctly
described by this model, we still believe that
it has some use in illustrating general properties.
Furthermore, since the integral of $E_q$ are
constrained by known anomalous magnetic moments,
it should be clear that the overall scale of
any resulting effect is model independent --- 
even if details require a better description for
the GPDs.

However, despite all these caveats, the 
qualitative picture of large $\perp$ distortions,
such as the ones depicted in Fig. \ref{fig:distort},
should be model independent. As has been emphasized
in Ref. \cite{ijmpa}, the $\perp$ center for 
each flavor ($\perp$ flavor dipole moment) is
related to the anomalous magnetic moment contribution
from that flavor and one thus finds that the typical
scale for $\perp$ distortions is on the order of
$0.2\, fm$. Although these considerations do not
constrain the $x$-dependence of the $\perp$
distortion, at least they determine the expected
typical size of the effect.

Furthermore, from the fact that the momentum 
asymmetry arises from corelating the impact 
parameter space asymmetry (Fig. \ref{fig:distort})
with the impulse (Fig. \ref{fig:fpanel}) it is
also clear that the sign of the asymmetry in the
nucleon is model independent: if the proton has
spin up and if one looks into the direction of
the momentum transfer then leading $u$-quark
will on average pick up a $\perp$ momentum
to the right, while leading $d$-quarks will be
deflected to the left.

\section{Angular and $x$ Dependence of the Asymmetry}
In the above model, the SSA arises when one
correlates the angular asymmetry of the impact
parameter dependent parton distribution with
the lensing function $\IT (\bT)$. For a transversely
polarized target polarized for example in the
$x$-direction, the transverse distortion
of the impact parameter dependent PDFs is
described by the transverse gradient of the
Fourier transform of $E(x,0,-\D2)$
\be
q(x,\bT) &=&
\int d^2\DT e^{-i\DT \bT} H(x,0,-\D2)\\
&+& \frac{i}{2M}\frac{\partial}{\partial b_y}
\int d^2\DT e^{-i\DT \bT} E(x,0,-\D2).\nonumber
\ee
Since $E(x,0,-\D2)$ is axially symmetric, one thus
finds that the transverse distortion of $q(x,\bT)$
exhibits a pure $\sin \phi$ angular dependence.
In general, since the PDFs become transversely
distorted, the lensing function $\IT (\bT)$ may
also exhibit an axial asymmetry. However, we 
expect this effect to be small and in a first 
approximation we may take $\IT (\bT)$ to be
axially symmetric.

In our simple description for the SSA, the final
state interaction is always directed toward the
center of the hadron. As a result, the final
state transverse momentum always point (anti-)
parallel to the transverse position $\bT$ that
the active quark had before it was struck
\be
{\cal P}(x,\kT) = \int d^2\bT q(x,\bT)
\delta\left(\kT - \IT (\bT)\right).
\label{eq:PxkT}
\ee
Therefore the pure pure $\sin \phi$ asymmetry of 
the PDFs in impact parameter space translates into a pure 
$\sin \phi$ asymmetry for the $\kT$ distribution of the SSA.

Because of the very large transverse distortion of
the impact parameter dependent PDFs entering Eq. 
(\ref{eq:PxkT}), it appears that our model 
predicts asymmetries that are numerically very large
($\sim 0.5$). However, one must keep in mind that
our description neglects (among other things) the
stochastic nature of the FSI, i.e. even in a
semi-classical model one would expect some smearing
of $\kT$, which would effectively reduce the
asymmetry. However, the fact that the asymmetry
depends only on $\sin \phi$ remains.

In order to determine the actual mean transverse 
momentum in our model, one needs to correlate
the transverse distortion (Fig. \ref{fig:distort})
with the $\perp$ impulse associated the impact 
parameter. Because there are significant 
uncertainties both in the choice of the vector
potential (and hence in $\IT(\bT)$) as well as in
the $x$-dependence of $E(x,0,t)$ (and hence in
$q(x,\bT)$) the resulting numerical values 
(in particular their $x$ dependence) would also
exhibit a large uncertainties. 

Nevertheless, we know that typical
$\perp$ distortions for $\perp$ polarized targets 
are on the order of $0.2\, fm$ 
(Fig. \ref{fig:distort}). And we also know
that the typical impulse acquired by an outgoing
quark for $\bT\approx 0.2\, fm$ is on the order of
$\IT \approx 0.3-0.5\, GeV/c$.
As a result, the natural scale for the transverse
momenta that emerges from this picture is also
on the order $\langle \kT\rangle 
\approx 0.2-0.4\, GeV/c$.
Generating such a large mean transverse momentum
scale in a natural way is one of the main results
of this paper.

\section{Summary}
We have provided a physical 
interpretation of the mechanisms that leads to a 
transverse single spin asymmetries (SSAs) in 
semi-inclusive electro-production of mesons. 
The starting point of our analysis is the 
BHS/Collins Wilson-line phase \cite{hwang,collins} 
that describes the final state interaction 
experienced by the active quark.
As the active quark escapes from the target, the 
chromodynamic gauge field from the remaining 
spectators provides an impulse that translates the
axial asymmetry in transverse position into an 
axial asymmetry in the transverse momentum of the 
outgoing quark before it fragments. This is the
physics that underlies the observation that
a Sivers asymmetry is allowed when one takes
final (or initial) state interactions into account.
Although the FSI are (to leading order) spin 
independent, they translate the spin-dependent
impact parameter space distributions into
a spin dependent transverse momentum of the leading
quark.

Because of the attractive nature of the confining 
interaction in QCD, the mean impulse on the 
outgoing quark is directed toward the center 
(of momentum) of the target.
The FSI with the spectators thus acts like a 
convex lens that deflects the active quark
toward the center.
The transverse impulse acquired by the active
quark is a direct consequence of the the Wilson 
phase factor advocated in Ref. \cite{hwang}.

In this simple picture, it is the combination
of the transverse position space asymmetry with 
this chromodynamic lensing effect that gives 
rise to the transverse momentum asymmetry
of the knocked out quarks.
Since the axial asymmetry of impact parameter 
dependent PDFs for transversely polarized nucleons 
tends to be rather large, this simple picture 
provides a natural mechanism for generating large
transverse single-spin asymmetries. 

Of course, our semi-classical picture cannot give 
an accurate description of the actual 
dynamics that gives rise to SSAs, but hopefully it 
will still help to provide a better understanding 
of the physics that leads to these asymmetries.

Acknowledgments: I appreciate discussions with  H. Avagyan,
N. Makins, and G. Schnell.
This work was supported by the Department of Energy 
(DE-FG03-96ER40965).
\appendix
\section{Detailed Derivation}
\label{sec:details}
First we use \cite{collins} the fact that time-reversal invariance
relates transverse momentum distributions with gauge strings that
point forward and backward in time respectively
\be
& &\\
\left. P(x,\kT,\sT)\right|_{future-pointing} &=&
\left. P(x,\kT,-\sT)\right|_{past-pointing} .
\nonumber
\ee
After performing a $180^o$ rotation around the $z-axis$ this implies
\be & &\\
\left. P(x,\kT,\sT)\right|_{future-pointing} &=&
\left. P(x,-\kT,\sT)\right|_{past-pointing}.
\nonumber
\ee
Upon evaluating the mean $\perp$ momentum, we thus find
\be
\PT(x,\sT)&\equiv& \left.\PT(x,\sT)\right|_{future}
\\
&=&-\left.\PT(x,\sT)\right|_{past}\nonumber\\
&=&\frac{1}{2}\left[ \left.\PT(x,\sT)\right|_{future}- 
\left.\PT(x,\sT)\right|_{past}\right].\nonumber
\ee
Using partial integration, we transform the integration in
$\left.\PT(x,\sT)\right|_{future/past}$ into the a $\perp$ derivative.
Using again the fact that only the $\perp$ derivative on the gauge 
field contributes, this implies after some straightforward algebra
\be
\PT(x,{\bf s_\perp}) &=& -\frac{g}{2}
\int_{-\infty}^\infty \frac{dy^- }{4\pi}
e^{-ixp^+y^-}\label{eq:Pi2}\\
& &\!\!\!\!\!\!\!\!\!\!\!\!\!\!\!\!\!\!\!\!\!\!\!\!\times
\left\langle p \left|\bar{q}(y^-) 
\int_{-\infty}^\infty dz^- W^\dagger_{yz} {\bf \partial_\perp}
A^+(z^-) W_{z0} \gamma^+ q(0)\right|p\right\rangle ,
\nonumber
\ee
where $\yT = \zT=\0T$, plus a term where the 
$\perp$ derivative acts on the quark and which
vanishes due to time reversal invariance.
Eq. (\ref{eq:Pi2}) is still rigorous.

Although the strict light-cone gauge $A^+=0$ 
is unattainable,
we conjecture that a gauge choice can be achieved where 
light-like gauge strings that do not extend to infinity become 
trivial. The motivation for this conjecture relies on a
limiting procedure, where works in a gauge $A^+=const.$ and imposes 
periodic boundary conditions in the $x^-$ direction. Upon taking the
`box-length' to infinity, all gauge strings of finite length become 
trivial
\be
W_{yz} \longrightarrow 1 \quad \quad \mbox{for}
\quad \quad y^-,z^-\not\in \{
-\infty,\infty\}. \label{eq:trivial}
\ee
This conjecture is consistent with a probabilistic 
interpretation for the twist-2 parton distributions probed 
in deep-inelastic scattering.

However, despite Eq. (\ref{eq:trivial}) we may not drop $A^+$ entirely
in Eq. (\ref{eq:Pi2}) since the integration over $z^-$ extends
to $\pm \infty$. In fact, the remaining contribution from the
gauge field is exactly the zero-mode advocated in 
Section
\ref{sec:main}

\bibliography{fsi5.bbl}
\end{document}